\documentclass[twocolumn,tighten]{aastex61}

\submitjournal{ApJ}

\usepackage[sort&compress]{natbib}
\bibpunct{(}{)}{;}{a}{}{,}
\usepackage{color}
\definecolor{darkgreen}{RGB}{0,142,128}

\usepackage{hyperref}
\hypersetup{colorlinks,citecolor=blue,linkcolor=blue}

\usepackage{amsmath}
\usepackage{amsthm}
\usepackage{amsfonts}
\usepackage{url}
\usepackage{subfigure}
\usepackage{multirow}

\newcommand{\gafd}{GAFD}

\shorttitle{The fate of close-in planets: tidal or magnetic migration?}
\shortauthors{A. Strugarek \textit{et al.}}

\begin{document}

\title{The fate of close-in planets: tidal or magnetic migration?}


\correspondingauthor{Antoine Strugarek}
\email{antoine.strugarek@cea.fr}

\author{A. Strugarek}
\affil{Laboratoire AIM Paris-Saclay, CEA/Irfu Universit\'e Paris-Diderot CNRS/INSU, F-91191 Gif-sur-Yvette.}

\author{E. Bolmont}
\affil{Laboratoire AIM Paris-Saclay, CEA/Irfu Universit\'e Paris-Diderot CNRS/INSU, F-91191 Gif-sur-Yvette.}

\author{S. Mathis}
\affil{Laboratoire AIM Paris-Saclay, CEA/Irfu Universit\'e Paris-Diderot CNRS/INSU, F-91191 Gif-sur-Yvette.}

\author{A. S. Brun}
\affil{Laboratoire AIM Paris-Saclay, CEA/Irfu Universit\'e Paris-Diderot CNRS/INSU, F-91191 Gif-sur-Yvette.}

\author{V. R\'eville}
\affil{UCLA Department of Earth, Planetary and Space Sciences, 595 Charles E. Young Dr East, Los Angeles CA 90095}
\affil{Laboratoire AIM Paris-Saclay, CEA/Irfu Universit\'e Paris-Diderot CNRS/INSU, F-91191 Gif-sur-Yvette.}

\author{F. Gallet}
\affil{Department of Astronomy, University of Geneva, Chemin des Maillettes 51, 1290 Versoix, Switzerland}

\author{C. Charbonnel}
\affil{Department of Astronomy, University of Geneva, Chemin des Maillettes 51, 1290 Versoix, Switzerland}
\affil{IRAP, UMR 5277, CNRS and Universit\'e de Toulouse, 14, av. E. Belin, F-31400 Toulouse, France}


\begin{abstract}
Planets in close-in orbits interact magnetically and tidally with their host stars. These interactions lead to a net torque that makes close-in planets migrate inward or outward depending on their orbital distance. We compare systematically the strength of magnetic and tidal torques 
for typical observed star-planet systems (T-Tauri \& hot Jupiter, M dwarf \& Earth-like planet, K star \& hot Jupiter) 
based on state-of-the-art scaling-laws. 
We find that depending on the characteristics of the system, tidal or magnetic effects can dominate. For very close-in planets, we find that both torques can make a planet migrate on a timescale as small as 10 to 100 thousands of years. Both effects thus have to be taken into account when predicting 
the evolution of compact systems.
\end{abstract}



\section{Introduction}
\label{sec:introduction}


Thanks to space missions such as CoRoT \citep{Corot}, {\it Kepler} \citep{Borucki:2010dn} and K2 \citep{Howell:2014ju} and ground-based observations (e.g. HARPS, \citealt{Pepe:2000ej}), about 3000 exoplanets have been discovered as of today since the pioneering detection by \citet{Mayor:1995cp}.
The corresponding planetary systems are very diverse in terms of planetary size and mass as well as orbital architecture. 
Due to the observational biases of the two most prolific detection techniques (transit and radial velocity), a majority of the 
detected exoplanets are close-in planets, which are very likely strongly interacting with their host star.

\begin{figure}[htb]
  \centering
  \includegraphics[width=\linewidth]{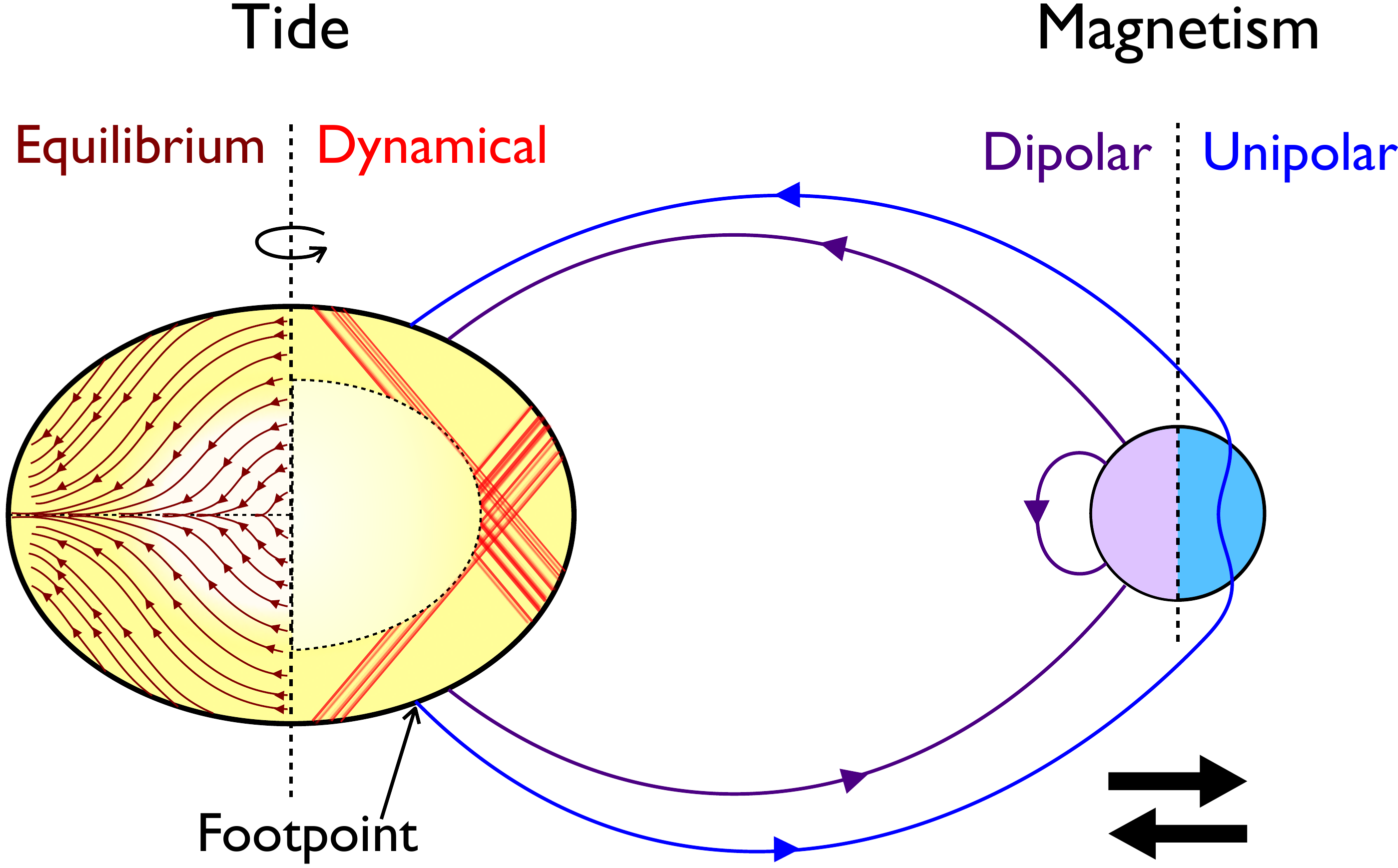}
  \caption{Sketch of the star-planet tidal and magnetic interactions. Only the tides (equilibrium tide and dynamical tide in the convective envelope) raised by the planet in the star are here taken into account. 
  We consider both dipolar and unipolar magnetic interactions. Due to these interactions, the planet can migrate either inward or outward (black arrows).}
  \label{fig:schema}
\end{figure}

Star-planet interactions were proposed to have various effects on the dynamics and evolution of compact systems \citep{Cuntz:2000ef}, among which angular momentum exchanges
between the planet's orbit and the stellar spin (and to a lesser extent between the planet's orbit and the planet's spin).
We concentrate here on the former and we consider only planetary systems with one planet and no protoplanetary disk. 
These exchanges lead to the spin-up or spin-down of the star and the orbital migration of the planet due to two main physical processes: tidal and magnetic interactions (see Fig. \ref{fig:schema}).

Tidal interactions consist in the gravitational response of a given body (here, the star) to a perturber (here, the planet) and its effect on rotation and orbit. 
There are two different components of the response: the hydrostatic non wave-like equilibrium tide \citep[e.g.][]{Zahn:1966tw} and the dynamical tide. The dynamical tide can develop either in the radiative core of the star \citep[e.g.][]{Zahn:1975vr} or in its rotating convective envelope \citep[e.g.][]{Ogilvie:2007kt}. 
The resulting dissipation can be several orders of magnitude higher than the dissipation due to the equilibrium tide \citep{Ogilvie:2007kt,Bolmont:2016kp}, leading to a much faster orbital migration, especially for stars on the Pre-Main Sequence \citep{Bolmont:2016kp,Gallet:2017vx}.
Recent works on star-planet tidal interactions allowed to give estimates of the dissipation due to the dynamical tide in the convective region of a star of a given mass, age, metallicity and rotation \citep{Gallet:2017vx,Bolmont:2017uj}. 

Simultaneously, magnetic interactions develop due to the differential motion between the 
planet and the magnetized ambient stellar wind at the planetary orbit. A net magnetic torque applies to the planet, effectively transferring angular momentum between the planet and star (if the planet is in the sub-alfv\'enic region of the wind, close to the star), or between the planet and the ambient wind (if the planet is in the super-alv\'enic region of the wind). In the context of close-in planets, we consider here only the former interaction. Different regimes of the interaction occur depending on the magnetic properties of the planet \citep[e.g.][]{Zarka:2007fo,Strugarek:2014gr}. If the planet possesses an intrinsic magnetic field, the \textit{dipolar} interaction develops \citep{Saur:2013dc,Strugarek:2015cm,Strugarek:2016ee}, otherwise the interaction becomes \textit{unipolar} \citep{Laine:2012jt}. The magnetic interaction between a star and a close-in planet can lead to many other notable effects such as anomalous emissions or planet inflation. A recent review on those effects can be found in \citet{Lanza:2017vy} and references therein.

In past studies about star-planet interactions, tides and magnetism have not been taken into account together, with the notable exception of \citet{Bouvier:2015kq} in the context of young fast rotators.
The aim of this letter is thus to offer the first generic comparison of the strength of tidally-induced versus magnetically-induced orbital migration for typical compact star-planet systems. As a first step, we focus here on planets on a circular coplanar orbit.
In \S~\ref{sec:estim-torq-appl}, we estimate the migration timescales due to the tidal and the magnetic interactions.
In \S~\ref{sec:migr-time-scal}, we compute the two migration timescales for 3 representative stars of different rotation and hosting planets of different orbital periods. We establish the parameter space for which tides or magnetism is the strongest driver of migration, and give the example of three particular observed exo-planetary systems.

\section{Estimates of torques applied to close-in planets}
\label{sec:estim-torq-appl}

We characterize the orbital migration of planets by the timescale defined as 
\begin{equation}
  \label{eq:MigTimeScale}
  \tau_{\rm mig} = \left|\frac{a}{\dot{a}}\right| = \frac{2J_P}{|\Gamma|}\, ,
\end{equation}
where $\Gamma$ is the torque applied to the orbiting planet,
$J_p = M_p\sqrt{G M_\star a}$ is the orbital angular momentum, $M_p$ and $M_\star$ the masses of the planet and the star, and $a$
is the semi-major axis of the planet. 

\subsection{Tidal Torque}
\label{sec:tidal-torque}

The torque associated with tidal dissipation in the host star leads to planet migration that scales as \citep{Kaula:1964ex,Jackson:2008ip}
\begin{equation}
  \label{eq:TorqueTides}
  \left| \Gamma_T \right| = 6 J_p 
  n \frac{M_p}{M_\star}
  \left(\frac{R_\star}{a}\right)^5
  \frac{k_2}{Q_*}\, ,
\end{equation}
where $n=\sqrt{\frac{GM_\star}{a^3}}$ is the orbital frequency, $R_\star$ the stellar radius, $k_2$ the usual quadrupolar hydrostatic Love number and $Q_{*}$ the tidal dissipation quality factor.

Planets induce tidal flows in stars that have two components: the equilibrium and the dynamical tides. The equilibrium tide is a large-scale non wave-like flow \citep[e.g.,][and Fig.~\ref{fig:schema}]{Zahn:1966tw,Remus:2012ef} sustained by the hydrostatic adjustment of the star because of the perturbation induced by a close planet. It is efficiently dissipated in the turbulent convective envelope of low-mass stars while its damping can be neglected elsewhere \citep{Zahn:1977wz}. However this flow is not a complete solution of the hydrodynamics equations and it is completed by the so-called dynamical tide \citep{Zahn:1975vr}. In the convective envelope of late-type stars, the dynamical tide is constituted by inertial waves governed by the Coriolis acceleration and damped by the convective turbulent friction as in the case of the equilibrium tide \citep{Ogilvie:2007kt}. They are excited only when $n < 2 \Omega_\star$ ($\Omega_\star$ is the stellar rotation rate). \cite{Mathis:2015ba} demonstrated that their dissipation is efficient when a radiative core is present for sufficiently thick convective envelopes. This allows the formation of sheared wave attractors and an important enough volume where dissipation can take place. In fully convective stars, for which attractors cannot form in the case of rigid rotation, and for $n > 2 \Omega_\star $, only the equilibrium tide prevails. 
The dynamical tide can also occur in the radiative core \citep{Zahn:1975vr,Ogilvie:2007kt}, but in this first work, we will focus on the dissipation in the convective envelope only. We thus need an accurate estimate of $k_2/Q_\star$ for each type of tide.


\subsubsection{Equilibrium Tide}
\label{sec:eqtide}

The dissipation of the equilibrium tide can be estimated using the analytical
model developed by \citet{Remus:2012ef}. In this model, the tidal
dissipation is written as
\begin{equation}
  \label{eq:q_eq_phys}
  \frac{k_2}{Q_*} = 4\pi\frac{2088}{35} \frac{R_\star^4}{G M_\star^2}
  \left| \sigma \int_{\alpha}^{1} x^8 \rho \nu_t \, {\rm d}x \right|\, ,
\end{equation}
where $\sigma = 2\left(n-\Omega_\star\right)$ is the tidal frequency in the coplanar circular case studied here, $\nu_t$ the turbulent viscosity in the convection zone, $\alpha = R_{bcz}/R_\star$, $R_{bcz}$ being the radius of its base, $\rho$ the density in the convection zone, and $x=r/R_\star$ the normalized radial coordinate. We derive hereafter a estimate of Eq. \eqref{eq:q_eq_phys} based only on the global parameters of the system.

The turbulent viscosity strength depends on how the convective turnover time $t_c$ compares to the tidal frequency $\sigma$. Following \citet{Zahn:1966tw} and \citet{Remus:2012ef}, we can generically write 
\begin{equation}
  \label{eq:turb_regimes}
  \nu_t = \frac{1}{3}v_c l_{c} \left[1+\left(\frac{t_{c}\sigma}{\pi}\right)^2 \right]^{-1/2}\, ,
\end{equation}
where $v_c$ is the typical convective velocity and $l_{c}$ the mixing
length. These three
parameters can be estimated using the derivation from \citet{Mathis:2016gy} based on the mixing length theory for a rotating body (proposed in \citealt{Stevenson:1979fq} and validated by recent high resolution numerical simulations \citealt{Barker:2014cn}). We define the convective Rossby number as $R_o^c=t_\Omega/t_c^0$ ($t_\Omega$ being the rotation period of
the star and $t_{c}^{0}$ the convective turnover time from the standard mixing length theory that neglects rotation), and use the following estimates \citep{Mathis:2016gy}
\begin{eqnarray}
  \label{eq:Vc}
  v_c &\simeq& \left(\frac{L_\star}{\bar{\rho}_{CZ}R_\star^2}\right)^{1/3} \!\!\!\!\!\!\times \left\{ 
  \begin{array}{cc}
    \left(1 - \displaystyle{\frac{1}{242 \left(R_o^c\right)^2}}\right) & \mbox{if } R_o^c > 0.25 \\
    1.5 \left(R_o^c\right)^{1/5} & \mbox{if
                                                             } R_o^c < 0.25
  \end{array}
  \right. \, , \\
  \label{eq:lc}
  l_c &\simeq& \alpha_{\rm MLT} H_P \times \left\{ 
  \begin{array}{cc}
    \left(1 + \displaystyle{\frac{1}{82 \left(R_o^c\right)^2}}\right)^{-1}  & \mbox{if } R_o^c > 0.25\\
    2 \left(R_o^c\right)^{3/5} & \mbox{if } R_o^c < 0.25
  \end{array}
  \right. \,  , \\
  \label{eq:tc}
  t_c 
  &=& \frac{l_c}{v_c} 
          \, ,
\end{eqnarray}
where we have introduced the stellar luminosity $L_\star$, the average
density in the convection zone $\bar{\rho}_{CZ}$, the mixing-length
parameter $\alpha_{\rm MLT}$, and the pressure scale height $H_P$. Given that we aim here for 
an order of magnitude estimate, we perform the further approximations that $\rho=\bar{\rho}_{CZ}$ and that $\nu_t$ in the integral in Eq.~\eqref{eq:q_eq_phys} does not vary with depth. Furthermore, we approximate the mixing length $l_{c}$ 
by its maximum which is given by the depth of the convection zone $(1-\alpha)R_\star$. We set $\bar{\rho}_{CZ} = 3 M_\star\left(1-\beta\right)/4\pi R_\star^3\left(1-\alpha^3\right)$ to the density of the convective envelope, with $\beta=M_{bcz}/M_{*}$ ($M_{bcz}$ is the radiative core mass). We finally obtain 
\begin{equation}
  \label{eq:dissip_end}
  \frac{k_2}{Q_*} \simeq \frac{232}{35} \frac{\left| t_c\sigma \right|}{\sqrt{1+\left(\displaystyle{\frac{t_{c}\sigma}{\pi}}\right)^2}} \frac{R_\star}{G M_\star}v_c^2 (1-\beta)\frac{1-\alpha^9}{1-\alpha^3}  \, 
  .
\end{equation}

\subsubsection{Dynamical Tide: Tidal Inertial Waves}
\label{sec:dyntide} 

The case of tidal inertial waves excited in the convective envelope of low-mass stars is complex to treat. As demonstrated by \citet{Ogilvie:2007kt}, the induced tidal torque strongly depends on the tidal frequency and can vary over several orders of magnitude as a function of the stellar mass, age, metallicity, rotation, and turbulent viscosity. 
Hence, a coherent treatment of this tidal dissipation requires the coupling of hydrodynamical numerical codes to compute tidal inertial waves \citep[e.g.][]{Guenel:2016di}, and 
rotational and orbital evolution codes to take into account their dissipation along the evolution planetary systems \citep[e.g.][]{Bolmont:2015ge,Bolmont:2016kp}. An alternative approach has been proposed by \citet{Ogilvie:2013bk}, \citet{Mathis:2015ba} and \citet{2015sf2a.conf..401M} who estimated the order of magnitude of the friction induced by the dissipation of tidal inertial waves thanks to analytical frequency-averaged dissipation rates derived using spherical bi-layer rotating stellar structure models. They obtain: 
\begin{eqnarray}
\lefteqn{  \frac{k_2}{Q_*} =                          \frac{100\pi}{63}\epsilon^2\frac{\alpha^5\left(1-\gamma\right)^2}{1-\alpha^5}}
  \nonumber \\ &\times& 
  \frac{\left(1-\alpha\right)^4\left(1+2\alpha+3\alpha^2+\frac{3}{2}\alpha^3\right)^2
                \left(1+\frac{1-\gamma}{\gamma}\alpha^3\right)}{\left[1+\frac{3}{2}\gamma+\frac{5}{2\gamma}\left(1+\frac{1}{2}\gamma-\frac{3}{2}\gamma^2\right)\alpha^3-\frac{9}{4}\left(1-\gamma\right)\alpha^5\right]^{2}}
                               \nonumber \, ,\\
  \label{eq:qprimeexpression}
\end{eqnarray}
where $\gamma=\displaystyle{\frac{\alpha^3\left(1-\beta\right)}{\beta\left(1-\alpha^3\right)}}$ and $\epsilon = \Omega_\star/\Omega_{\rm kep}(R_\star)$ with the keplerian rotation rate $\Omega_{\rm kep}(r) = \sqrt{GM_\star/r^3}$. In this work, we use the stellar evolution grid of \citet{Gallet:2017vx}\footnote{See \href{https://obswww.unige.ch/Recherche/evol/starevol/Galletetal17.php}{this link}} to estimate $\alpha$ and $\beta$ for given star.

\subsection{Magnetic Torque}
\label{sec:magnetic-torque}

\subsubsection{Stellar Wind Model}
\label{sec:stellar-wind-model}

The magnetic interaction in compact star-planet systems depends on the local properties of the stellar wind at the planetary orbit. In order to estimate these properties, we use the simple 1D magnetized wind model \textit{starAML} (see \citealt{Reville:2015bb} for the complete description, the code is available upon request to the authors). 
This model solves the
pressure balance of a magnetized 1D Parker-like wind
\citep{Parker:1958dn,Weber:1967kx,Sakurai:1985uc}. The magnetic field of the wind is extrapolated with a potential field with a source surface \citep{Schatten:1969cn} calibrated over 2D and 3D numerical simulations of stellar winds \citep[see][for 3D]{Reville:2016hw}.  
The general properties of the wind such as velocity, mass and angular momentum loss rates, and Alfv\'en radius are then simply computed. 
In the present work we assume a ratio of specific heats $c_p/c_v = 1.05$, which mimics the coronal heating and allows to reproduce the solar wind with this model. We scale the  density and temperature at the base of the corona with the rotation rate of the star using the observational prescription of \citet{Holzwarth:2007hl}. 
The large-scale magnetic field of the star $B_s$ is assumed to depend on the rotation rate of the star as
\begin{equation}
B_s = B_\star \left\{
\begin{array}{cc}
1 & \mbox{if } R_o^c < 0.25 \\
\left(\frac{R_o^c}{0.25}\right)^{-1.38} & \mbox{if } R_o^c > 0.25 
\end{array} \right. \, ,
\end{equation}
where the exponent $-1.38$ is taken from the empirical trend found by \citet{Vidotto:2014ba}, and $B_\star$ is the reference stellar field in the rotationally-saturated regime ($R_o^c < 0.25$). We choose here a threshold at $R_o^c=0.25$ to be consistent with the threshold for tides (see \S \ref{sec:eqtide}), which is close to the observationally-constrained saturation value \citep[e.g.][]{Gondoin:2012ct,Vidotto:2014ba}. 

With this wind model, we can furthermore infer whether the planet is able to sustain a magnetosphere for a given planetary magnetic field surface amplitude $B_p$. In this work we consider a dipolar or an unipolar regime of interaction depending whether a magnetosphere can be sustained or not (see Fig. \ref{fig:schema}).

\subsubsection{Unipolar Interaction}
\label{sec:unipolarInteraction}

In cases where the ambient pressure from the stellar wind is too strong, the planet magnetic field may not be strong enough to sustain a magnetosphere. In this case, the magnetic interaction in a close-in system becomes unipolar and was studied by \citet{Laine:2008dx,Laine:2012jt}. The magnetic torque can then be written as
\begin{equation}
  \label{eq:LaineLinUnipolar}
  \left| \Gamma_M \right| = 8 R_p^2 a^2 \left|\sigma\right| B_w \Sigma \, ,
\end{equation}
where $R_p$ is the planet radius, $B_w$ is the stellar wind magnetic field at the planetary orbit and $\Sigma$ is the resistance of the stellar plasma at the footpoint of the interaction (see Fig. \ref{fig:schema}) that was estimated to be of the order of $6.5\,\times 10^{-6}$ s/cm by \citet{Laine:2012jt}. Note that we have simplified the original equation to neglect the dependency of the torque to the incidence angle of the interaction footpoint at the stellar surface, for the sake of simplicity. The torque \eqref{eq:LaineLinUnipolar} was derived assuming a rocky super-Earth planet in \citet{Laine:2012jt}, and is valid as long as ohmic dissipation is more efficient at the footpoint of the interaction at the stellar surface than inside the planet itself. This assumption may also hold for gaseous planets, depending on their poorly constrained conductivity profile in their interior \citep[e.g.][]{Umemoto:2006gna,vandenBerg:2010bk}. We suppose here that Eq. \eqref{eq:LaineLinUnipolar} holds for all planets considered, but warn the reader that the unipolar torque may decrease if the conductivity inside the planet is comparable to or less than the conductivity at the stellar surface.

\begin{deluxetable*}{lccccccccccl}[t]
\tablecaption{Representative systems\label{ta:repsys}}
\tablecolumns{11}
\tabletypesize{\scriptsize}
\tablehead{
\colhead{Star-planet system} &
\colhead{$M_\star$ [$M_\odot$]} &
\colhead{$R_\star$ [$R_\odot$]} &
\colhead{$B_\star$ [G]} & 
\colhead{$\alpha^a$} & 
\colhead{$\beta^a$} & 
\colhead{$T_{\rm eff}^a$ [K]} & 
\colhead{$M_p$ [$M_\oplus$]} & 
\colhead{$R_p$ [$R_\oplus$]} & 
\colhead{$P_{\rm orb}$ [days]} &
\colhead{$B_p$ [G]} &
\colhead{Analog system}
}
\startdata 
    T-Tauri + hot Jupiter & 1.05 & 1.12 & 2500 & 0.583 & 0.728 & 4475 & 318 & 11 & 10.9 & 10 & Tap-26 (b)$^b$ \\
    M dwarf + Earth & 0.13 & 0.17 & 1000 & 0 & 0 & 3070 & 0.3 & 0.7 & 0.45 & 1 & Kepler-42 (c)$^c$\\
    K star + hot Jupiter & 0.80 &  0.80 & 140 & 0.672 & 0.929 & 4875 & 363 & 12.5 & 2.2 & 28$^{e}$ & HD 189733 (b)$^d$ \\
\enddata
\tablecomments{The stellar parameters ($a$) are taken from \citet{Gallet:2017vx}. The planet parameters are derived from \citet{Yu:2017cc} ($b$), \citet{Muirhead:2012dx} ($c$) and \citet{Bouchy:2005kv} ($d$). The magnetic field of the star and the planet are assumed here, except for the third system ($e$) for which we took the value inferred by \citet{Cauley:2015kl} using observed abnormal pre-transit absorptions to estimate the size of the planet's hypothetical magnetosphere.}
\end{deluxetable*}

\subsubsection{Dipolar interaction}
\label{sec:dipolarinteraction}

The torque $\Gamma_M$ associated with magnetic forces has been modeled in
\citet{Strugarek:2015cm} and \citet{Strugarek:2016ee} when the planet possesses a magnetosphere (the so-called dipolar regime). It can be parametrized as
\begin{equation}
  \label{eq:TorqueMag}
  \left|\Gamma_M\right| = A_0 \pi \left(c_d P_t
    M_a^\chi \right) \cdot \left(R_p^2 a \right) \Lambda_p^\xi \, ,
\end{equation}
where $P_t$ is the stellar wind total pressure at the planetary orbit, $M_a$
the alfv\'enic Mach number, and $\Lambda_p$ the pressure ratio between
the magnetic pressure in the 
magnetosphere of the planet and the wind pressure $P_t$. The
coefficients $A_0$, $\chi$ and $\xi$ have been calibrated from a set of
3D numerical simulations in \citet{Strugarek:2016ee}, and depend on the
magnetic topology of the interaction. We will consider in this letter
only the aligned configuration, which maximizes the magnetic
torque. Note that we also omitted the dependency of the torque to the
resistive properties of the wind plasma, for the sake of simplicity
(see \citealt{Strugarek:2016ee} for a complete discussion). Finally, we assume that the planet magnetic field $B_p$ is independent of the spin of the planet (and hence of its orbital period in tidally-locked systems). This is a reasonable approximation as we expect the magnetic moment of a planet to depend only weakly on its rotation rate \citep{Christensen:2010eb,Davidson:2013gx}.

\section{Application to Particular Star-Planet Systems}
\label{sec:migr-time-scal}

Our goal is to compare the instantaneous timescales associated with tidal and magnetic forces (Eq. \ref{eq:MigTimeScale}). We thus define their ratio as 
\begin{equation}
	\Xi = \frac{\tau_{T}}{\tau_{M}}\, ,
\end{equation}
where the tidal ($\tau_{T}$) and magnetic ($\tau_{M}$) migration timescales are defined by Eq. \eqref{eq:MigTimeScale} with the torque $\Gamma_T$ defined by \eqref{eq:TorqueTides} for tides, and $\Gamma_M$ by \eqref{eq:LaineLinUnipolar} or \eqref{eq:TorqueMag} for magnetism. The overall migration timescale due to the sum of tidal and magnetic torques $\Gamma_T + \Gamma_M$ can be written as
\begin{equation}
\label{eq:TotalTime}
\tau = \frac{\tau_T\tau_M}{\tau_T+\tau_M}\, .
\end{equation}

The migration timescale formulae in \S~\ref{sec:estim-torq-appl} are generic to star-planet close-in systems. We now apply them to three illustrative systems listed in Table~\ref{ta:repsys}. 

\begin{figure*}[t]
  \centering
  \includegraphics[width=\linewidth]{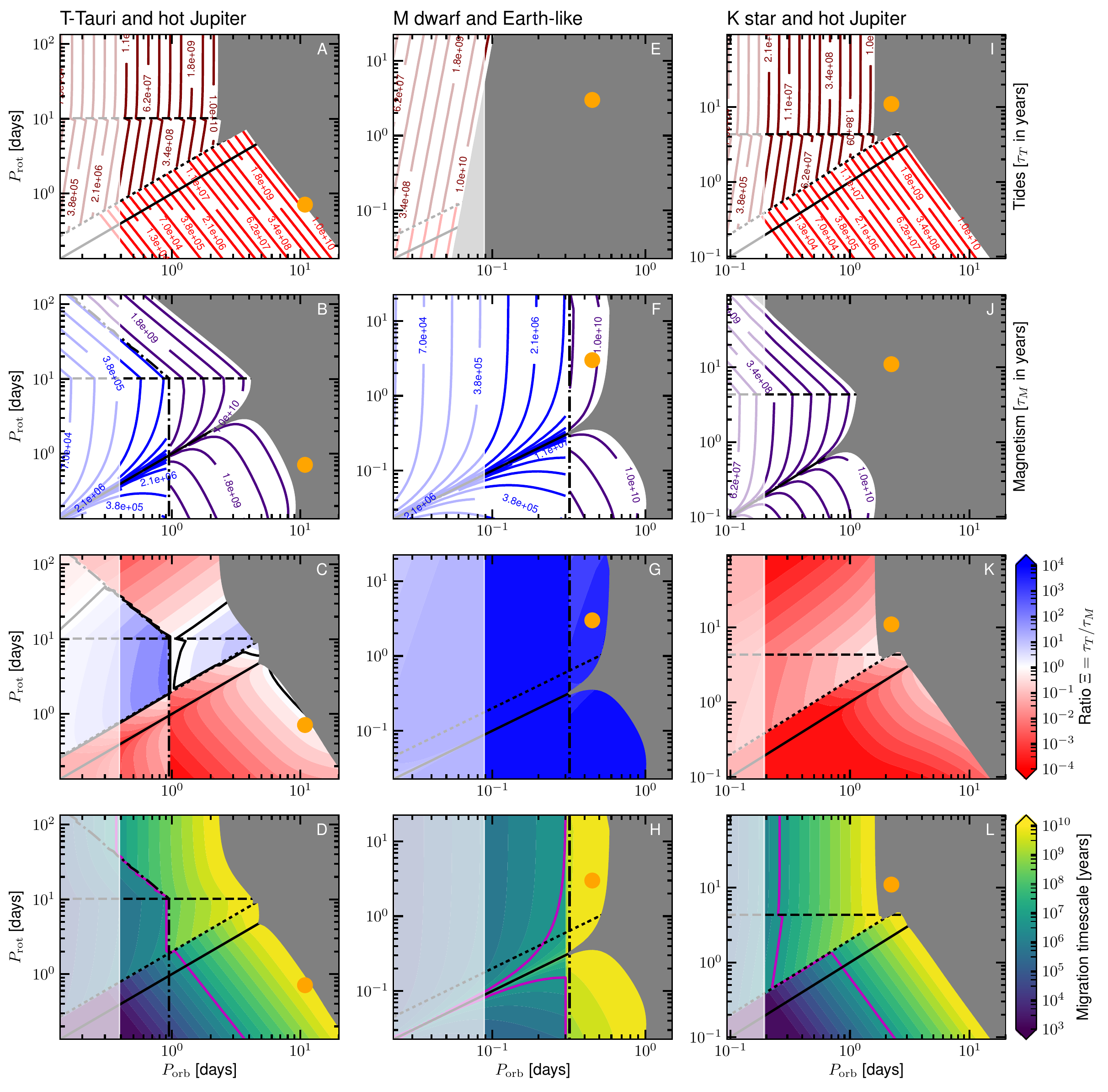}
  \caption{Tidal and magnetic torques in three representative star-planet systems (each column represents one system). The tidal (first row A, E, I) and magnetic (second row B, F, J) migration timescales are contoured with labels in years. The ratio between the tidal and magnetic migration timescales is shown in the third row (C, G, K), with blue (red) denoting magnetic (tidal) dominance. The overall migration time-scale is displayed in logarithmic scale in the last row (D, H, L). The magenta line corresponds to a timescale of 5~Myr, which is representative of a stellar structure evolution timescale on the Pre-Main Sequence. All panels are shown as a function of the rotation period of the star and the orbital period of the planet, in days. The gray area masks regions where the migration timescale is larger than the age of the universe. The Roche limit is labeled by the white transparent area and calculated as in \citet{Strugarek:2014gr}. The orange circles label the position of the three analog systems identified in Table \ref{ta:repsys}. The oblique black lines label co-rotation (plain line) and dynamical tide transition (dashed line). The horizontal black line with long dashes labels the transition at $R_o^c=0.25$, and the dash-dotted line traces the change of magnetic interaction regime.}
  \label{fig:FinalFig}
\end{figure*}

We first consider the case of a T-Tauri star like Tap 26 \citep{Yu:2017cc}, with a surrounding hot Jupiter possessing a magnetic field twice stronger than Jupiter's. The T-Tauri is a young solar-like star, that is assumed here to generate a strong surface magnetic field of the order of $2500$ G \citep[e.g.][]{Donati:2009if}. The resulting tidal and magnetic migration timescales are shown in the left column of Fig. \ref{fig:FinalFig} as a function of the rotation period of the star and the orbital period of the planet (the Tap 26 system is labeled by the orange circle). 

The iso-contours of the tidal migration timescale are shown in the first panel (A). The grey area masks regions where the torque is too weak and the associated migration timescale is larger than the age of the universe. The different regimes of the tidal migration efficiency clearly appear in this panel. A sharp transition is observed at $n=2\Omega_\star$ (oblique black dashed line): the dynamical tide (red) operates only in star-planet systems below this transition line, dramatically improving the efficiency of the tidal dissipation. A second transition is observed in the equilibrium tide region above (dark red), labeled by the horizontal black line with long dashes, where the star enters a large Rossby number regime (see \S~\ref{sec:eqtide}). 
In the upper part of the diagram, the tidal migration timescale becomes essentially independent of the slow rotation rate of the star, as expected.
The black oblique line represents the co-rotation radius: a planet below this line migrates outward and planet above this line migrates inward.

The magnetic migration timescale is shown in the second panel (B). 
Because the Alfv\'en surface of the wind is located farther than a 10-day orbit, the wind is assumed here in co-rotation with the star, meaning that the co-rotation line is the same here as for tides.
We suppose that the planet possesses an intrinsic surface magnetic field of $10$ G. 
A sharp transition occurs at 0.9 days (black dash-dotted line) as the magnetic interaction changes from dipolar (dark blue) to unipolar (blue). The latter is shown to be more efficient than the effective drag occurring in the dipolar interaction case.

The ratio between the two timescales $\Xi$ is shown in the third panel (C). The thick black contour labels the positions where the tidal and magnetic torques have the same amplitude. Red (blue) regions denote regions where the tidal (magnetic) torque dominates. 
We note that generally, tidal effects dominate in the regions where dynamical tides operate. For slowly rotating systems, though, both torques are clearly comparable with one slightly dominating the other depending on the orbital period of the planet. 

Finally, the overall migration timescale $\tau$ is shown in logarithmic scale in the fourth panel (D). The closest planets migrate on a timescale of several thousands of years due to both tidal and magnetic torques for most stellar rotation rates. Both torques strongly decrease with the orbital distance. 
As a result, planets with orbital periods longer than 10 days are essentially insensitive to tidal and magnetic torques.

The second column in Fig.~\ref{fig:FinalFig} (panels E-H) shows the same quantities for a fully-convective M-dwarf resembling Kepler-42 (\citealt{Muirhead:2012dx}, the orange circle labels Kepler-42~c). In this case the dynamical tide is much less efficient as there are no attractors, and as a result the tidal migration timescale is long for all rotation rates. The torque applied to close-in planets is completely dominated by magnetic effects, due to both the weak tidal torque and the relatively strong magnetic field of the M-dwarf. The critical period at which the star reaches a Rossby number of 0.25 
is 28 days in this case. 
Because of the relatively weak magnetic field of the planet, the magnetic interaction mostly develops in an unipolar regime in such systems. 
Only planets with orbital periods smaller than $\sim$0.3 days are expected to significantly migrate.

Finally, we study in the third column (panels I-L) the case of an evolved K star with a close-in hot Jupiter alike HD~189733 (\citealt{Bouchy:2005kv}, the orange circle labels HD~189733~b). The magnetic field of the star is much weaker than in the two previous cases, and the tidal torque completely dominates the migration path of the close-in planet (panel K). 
We immediately see that for the HD~189733 system, it is unlikely that the tidal and magnetic torques studied in this work can make the planet migrate significantly. Indeed, only planets with orbital periods smaller than $\sim$2 days are expected to migrate in such a system.


\section{Discussion and Conclusions}
\label{sec:discussion}

Magnetic and tidal effects generally act together in close-in systems to make planets migrate inward or outward. 
We found that both effects can dominate depending on the star-planet system considered. Magnetic effects are likely to dominate when the dynamical tide is not operating (e.g. for fully convective stars), and when the stellar magnetic field is strong. The overall torque depends on many parameters of the star-planet system, such as the stellar structure, its magnetic field and rotation rate; and the planet orbital distance, its structure, and its internal magnetic field. The multiple dependencies of the torques are fully accounted for in the simple scaling laws we summarized in this letter. When considering particular systems for which some parameters are not available, a simple parameter space exploration rapidly gives minimal and maximal migration timescale that can be attained due to tides and magnetic fields.


We have applied these scaling laws to three representative systems (Fig.~\ref{fig:FinalFig}). We found that very close-in planets could migrate on a timescale as small as 10 to 100 thousands of year due to the combination of tidal and magnetic interactions with their host. This migration timescale is short and renders the detection of such systems statistically unlikely. The observation of systems for which the magnetic or tidal torques are strong would provide a fantastic test-bed for the estimates we derived in this letter. For the three systems studied here, the known planets lie in a region of very long migration time-scale (orange disks). In the particular case of HD~189733, this suggests that no significant angular momentum transfer between HD~189733~b and its host is occurring due to either magnetic or tidal interactions. 

We have considered isolated, tidally-locked planets on a coplanar circular orbit. While this may be a reasonable assumption for single planet systems, it is not realistic for multiple planets systems. The tidal torque scaling-laws should be derived in the case of elliptic systems \citep[e.g. using][]{Kaula:1961hz}, but more theoretical work is still needed to develop estimates of the magnetic torques in such systems. The friction induced by the damping of tidal gravity waves in the radiative core of low-mass stars \citep{Zahn:1975vr} should also be taken into account in a near future, as well as tidal elliptic instabilities \citep{Cebron:2013jw}.

In the context of young star-planet systems, we have also neglected the contributions from interactions with a disk. Planet migration in a disk through Lindbald resonances is generally thought to be more efficient than the effects considered in this work \citep[e.g.][]{Baruteau:2014kn,Bouvier:2015kq}. We note though that the migration due to self-consistent dynamical tides was not systematically compared to disk-induced migration, which we leave for future work. 

Finally, we recall that we have focused our discussion on order of magnitude estimates of instantaneous migration timescales. The scaling-laws summarized in this letter could also be implemented in stellar and orbital evolution codes \citep{Zhang:2014iz,Bolmont:2015ge,Gallet:2017vx} to evolve self-consistently the star and its orbiting planet over secular time-scales, in particular when the migration timescale is longer than the characteristic evolution time of the star (see, e.g., the magenta line in Fig. \ref{fig:FinalFig}). For older stars such as the M dwarf and K star studied here, the stellar parameters are not expected to change significantly during their evolution over the Main Sequence, and our instantaneous estimates already provide a good approximation of the migration path of close-in planets.

As a general conclusion, stars and close-in planets are interacting both through gravitational and electromagnetic interactions and both of them should be taken into account to predict their evolution.

\acknowledgments

We thank D. C\'ebron and J. Bouvier for discussions about star-planet interactions and K. Augustson for discussions about scaling laws for stellar magnetic fields.
This work was supported by the \href{http://pnp-insu.fr}{Programme National de Plan\'etologie} (CNRS/INSU) and by a PLATO grant from the Centre National d'Etudes Spatiales (CNES) at DAp.
This work was also supported by the ANR 2011 Blanc 
\href{http://ipag.osug.fr/Anr\_Toupies/}{Toupies}
webmailand the ERC projects \href{http://www.stars2.eu/}{STARS2} (ERC grant 207430) and \href{http://irfu.cea.fr/Sap/en/Phocea/Vie_des_labos/Ast/ast_technique.php?id_ast=3914}{SPIRE} (ERC grant 647383). This work results within the collaboration of the COST Action TD 1308.


\end{document}